\documentclass{paper}
  \usepackage{amsfonts,amsmath,latexsym,amscd,graphicx,amssymb}
  \usepackage{paralist}
  \usepackage{graphics} 
  \usepackage{epsfig} 

  \usepackage{hyperref} 

\newtheorem{theorem}{Theorem}[section]

\newcommand{\be}{\begin{equation}}
\newcommand{\ee}{\end{equation}}
\newcommand{\ba}{\begin{eqnarray}}
\newcommand{\ea}{\end{eqnarray}}

\newcommand{\f}{\frac}

\title{Variational derivation of the Camassa-Holm shallow water
equation with non-zero vorticity }

\author{ \normalsize Delia IONESCU-KRUSE\\
\normalsize Institute of Mathematics of the Romanian Academy,\\
\normalsize P.O. Box 1-764, RO-014700, Bucharest,
 Romania\\
 \normalsize E-mail: Delia.Ionescu@imar.ro\\}

\begin{document}
\maketitle


\medskip

\bigskip


\begin{abstract}
 We describe the physical hypotheses underlying the
derivation of
 an approximate model of water waves.
 For unidirectional surface shallow water waves moving over
 an irrotational flow as well as over a non-zero vorticity flow,
 we derive the Camassa-Holm equation by an interplay of
 variational methods and small-parameter
expansions.
Subjclass: 35Q58, 76B15, 70G75.

\end{abstract}

\section{Introduction}
The Camassa-Holm equation reads

\begin{equation} U_t+2\kappa U_x+3UU_x-U_{txx}=2U_xU_{xx}+UU_{xxx}
\label{ch}\end{equation} with $x\in\mathbf{R}$, $t\in\mathbf{R}$,
$U(x,t)\in \mathbf{R}$. Subscripts here, and later, denote partial
derivatives. The constant $\kappa$ is related to the critical
shallow water speed.
 For
$\kappa$=0 the equation (\ref{ch}) possesses peaked soliton
solutions \cite{holm}. The physical derivation of (\ref{ch}) as a
model for the evolution of a shallow water layer under the
influence of gravity, is due to Camassa and Holm \cite{holm}. See
also Refs. \cite{adrian1},  \cite{dullin}, \cite{johnson1} for
alternative derivations within the shallow water regime. In
\cite{misiolek} it was shown that the equation (\ref{ch})
describes a geodesic flow on the one dimensional central extension
of the group of diffeomorphisms  of the circle (for the case
$\kappa$=0, a geodesic flow on the diffeomorphism group of the
circle, see also \cite{adrian&boris}). It should be mentioned
that, prior to Camassa and Holm, Fokas and Fuchssteiner \cite{ff}
obtained formally, by the method of recursion operators in the
context of hereditary symmetries, families of integrable equations
containing (\ref{ch}). These equations are bi-Hamiltonian
generalizations of the KdV equation and possess infinitely many
conserved quantities in involution \cite{ff}. In order to see how
the equation (\ref{ch}) relates with one of equations in the
families introduced in \cite{ff}, see \cite{f}, \S 2.2. Also, the
equation (5.3) in \cite{f2} is the equation (\ref{ch})  but with
errors in the coefficients.

The Camassa-Holm equation attracted a lot of interest, due to its
complete integrability \cite{ac1998}, \cite{CM}, \cite{gh},
\cite{v} (for the periodic case), \cite{BSS}, \cite{ac},
\cite{cgi} and the citations therein (for the integrability on the
line), the existence of waves that exist for all times as well as
of breaking waves \cite{CE} and the presence of peakon solutions
\cite{holm}. Of interest is also the fact that its solitary waves
are solitons, recovering their form and speed after nonlinear
interaction  \cite{BSS2}, \cite{johnson3}. These waves are smooth
if $\kappa \neq 0$ and peaked if $\kappa=0$, with both wave forms
being stable and thus physically recognizable \cite{cs1},
\cite{cs2}.
An important question to be answered is whether solutions of the
water wave problem can really be approximated by solutions of the
Camassa-Holm equation. In \cite{guido}  it is shown that suitable
solutions of the water wave problem and solutions of the
Camassa-Holm equation stay close together for long times, in the
case $\epsilon=\delta^2$, $\epsilon$ being the amplitude parameter
and $\delta$ the shallowness parameter.  In \cite{guido} it is
also shown that the peakon equation cannot strictly be derived
from the Euler equation and hence it is at most a phenomenological
model.

 In the references  \cite{dullin}, \cite{delia}, \cite{johnson1},
the derivation of the Camassa-Holm equation is made under the
assumption that the fluid motion is irrotational. This assumption
does not appear explicitly in the initial derivation \cite{holm}.
This derivation  is based on the Green-Naghdi equations and on the
choice of a submanifold in the Hamiltonian representation of the
Green-Naghdi equations. For the Green-Naghdi equations, one does
not have to impose any condition that the fluid is irrotational,
but, as shown in \cite{johnson1}, in order to have a consistency
for the second assumption one should impose an irrotational
condition on the flow. Thus, another important question to be
answered is the question whether a Camassa-Holm equation can be
obtained when the fluid below the free surface has non-zero
vorticity. By examining the underlying flow one has some important
and new consequences. In \cite{johnson2}  it is considered the
Camassa-Holm equation for water waves moving over a shear flow.
The calculation is completed for the case of a linear shear, that
is, the underlying flow has constant vorticity. In this case it is
shown that the equation for the surface wave is not a Camassa-Holm
equation, but the Camassa-Holm equation   can exist for  a simple
nonlinear function of the horizontal velocity component of the
perturbed flow field, at a certain depth.

The purpose of the present paper is to derive  by a variational
approach in the Lagrangian formalism, the Camassa-Holm equation
for unidirectional surface shallow water waves  when the fluid
below the free surface has non-zero vorticity.  We will consider
in turn the cases of an irrotational flow (for this case see also
\cite{delia}), a rotational flow with constant vorticity and
finally, an arbitrary flow. For the linear shear case,
 we show that the displacement of the free
surface from the flat state, satisfies the Camassa-Holm equation
(\ref{ch}) with $\kappa\neq 0$. For an arbitrary
 unidirectional shallow water flow,  we get that  the displacement
of the free surface from the flat state, satisfies a generalized
Camassa-Holm equation.

\section{Nondimensionalisation and scaling of the governing
equations for
 water waves}

We consider water moving in a domain with a free upper surface at
$z=h_0+\eta(x,t)$, for a constant $h_0>0$, and a flat bottom at
$z=0$.  The undisturbed water surface is $z=h_0$. Let $(u(x,z,t),
v(x,z,t))$ be the velocity of the water - no motion takes place in
the $y$-direction. The fluid is acted on only by the acceleration
of gravity $g$, and the effects of surface tension are ignored.
For  gravity water waves, the appropriate equations of motion are
Euler's equations (EE )(see \cite{johnson-carte}). Another
realistic assumption for gravity water wave problem is the
incompressibility (constant density $\rho$) (see
\cite{lighthill}), which implies the equation of mass conservation
(MC). The boundary conditions for the water wave problem are the
kinematic boundary conditions as well as the dynamic boundary
condition. The kinematic boundary conditions (KBC) express the
fact that the same particles always form the free water surface
and that the fluid is assumed to be bounded below by a hard
horizontal bed $z=0$. The dynamic boundary condition (DBC)
expresses the fact that on the free surface the pressure is equal
to the constant atmospheric pressure denoted $p_0$. Summing up,
the exact solution for the water-wave problem is given by the
system
\begin{equation}
\begin{array}{c}
\begin{array}{c}
u_t+uu_x+vu_z=-\f1{\rho} p_x\\  v_t+uv_x+vv_z=-\f1{\rho} p_z-g\\
\end{array}
\quad \quad \quad \quad \textrm{ (EE) }\\ \cr
 \qquad \qquad u_x+v_z=0  \qquad \qquad \qquad \qquad \textrm{ (MC)  }\\ \cr
\begin{array}{c}
  v=\eta_t+u\eta_x \, \, \textrm{ on }\,
z=h_0+\eta(x,t)\\
 v=0 \, \,
\textrm { on } z=0
\end{array}
\quad \textrm{ (KBC) }
\\ \cr
\qquad p=p_0, \,  \textrm{ on } z=h_0+\eta(x,t)
  \quad \qquad \textrm{ (DBC)} \end{array} \label{e+bc}
 \end{equation} where $p(x,z,t)$ denotes the pressure.
While a mathematical study of exact solutions   to the governing
equations (\ref{e+bc}) for water waves can be pursued (see for
example \cite{toland} for the periodic steady solutions in the
irrotational case and \cite{cs2'} for the periodic steady
solutions in the case of non-zero vorticity),  to reach detailed
information about qualitative features of water waves it is useful
to derive approximate models which are more amenable to an
in-depth analysis.

We non-dimensionalise the set of equations (\ref{e+bc}) using the
undisturbed depth of water $h_0$, as the vertical scale,  a
typical wavelength $\lambda$, as the horizontal scale, and a
typical amplitude of the surface wave $a$ (for more details see
\cite{johnson-carte}, \cite{johnson1}). An appropriate choice for
the scale of the horizontal component of the velocity is
$\sqrt{gh_0}$. Then, the corresponding time scale is
$\f\lambda{\sqrt{gh_0}}$ and the scale for the vertical component
of the velocity is $h_0\f{\sqrt{gh_0}}{\lambda}$. Thus, we define
the set of non-dimensional variables \begin{equation}
\begin{array}{c}
x\mapsto\lambda x,  \quad z\mapsto h_0 z, \quad \eta\mapsto a\eta,
\quad t\mapsto\f\lambda{\sqrt{gh_0}}t,\\
  u\mapsto  \sqrt{gh_0}u,
\quad v\mapsto h_0\f{\sqrt{gh_0}}{\lambda}v
\end{array} \label{nondim}\end{equation}
where, to avoid new notations, we have used the same symbols for
the non-dimensional variables  $x$, $z$, $\eta$, $t$, $u$, $v$, in
the right-hand side. The partial derivatives will be replaced by
\begin{equation}
\begin{array}{c}
u_t\mapsto \f{gh_0}{\lambda}u_t, \quad u_x\mapsto
\f{\sqrt{gh_0}}{\lambda}u_x, \quad u_z\mapsto\f {\sqrt{gh_0}}{h_0}u_z,\\

v_t\mapsto \f{gh_0^2}{\lambda^2}v_t, \quad v_x\mapsto
h_0\f{\sqrt{gh_0}}{\lambda^2}v_x, \quad v_z\mapsto\f {\sqrt{gh_0}}{\lambda}v_z,\\
\end{array}\label{derivate}\end{equation}

\noindent Let us now define the non-dimensional pressure. If the
water would be stationary, that is, $u\equiv v \equiv 0$, from the
first two equations and the last condition with $\eta=0$, of the
system (\ref{e+bc}), we get for a non-dimensionalised $z$, the
hydrostatic pressure $p_0+\rho g h_0(1-z)$. Thus, the
non-dimensional  pressure is defined  by \begin{equation} p\mapsto
p_0+\rho g h_0(1-z)+\rho g h_0 p \label{p}\end{equation} and
\begin{equation} p_x\mapsto \rho \f {gh_0}{\lambda} p_x, \quad
p_z\mapsto -\rho g+\rho g p_z\label{p'}\end{equation}

 Taking
into account (\ref{nondim}), (\ref{derivate}), (\ref{p}), and
(\ref{p'}), the water-wave problem (\ref{e+bc})  writes
 in
non-dimensional variables, as \begin{equation}
\begin{array}{c}
u_t+uu_x+vu_z=- p_x\\  \delta^2(v_t+uv_x+vv_z)=- p_z\\
 u_x+v_z=0\\
v=\epsilon(\eta_t+u\eta_x) \,   \textrm{ and } \,  p=\epsilon\eta
\, \, \textrm{ on }\,
z=1+\epsilon\eta(x,t)\\
 v=0 \, \,
\textrm { on } z=0
 \end{array}
\label{e+bc'} \end{equation}  where we have introduced the
amplitude parameter $\epsilon=\f a{h_0}$ and the shallowness
parameter $\delta=\f {h_0}{\lambda}$. The small-amplitude shallow
water is obtained in the limits $\epsilon\rightarrow 0$,
$\delta\rightarrow 0$. We observe that, on $z=1+\epsilon\eta$,
both $v$ and $p$ are proportional to $\epsilon$. This is
consistent with the fact that as $\epsilon\rightarrow 0$ we must
have $v\rightarrow 0$ and $p\rightarrow 0$, and it leads to the
following scaling of the non-dimensional variables
\begin{equation} p\mapsto \epsilon p,\quad
(u,v)\mapsto\epsilon(u,v) \label{scaling}\end{equation} where we
avoided again the introduction of a new notation. The problem
(\ref{e+bc'}) becomes \begin{equation}
\begin{array}{c}
u_t+\epsilon(uu_x+vu_z)=- p_x\\  \delta^2[v_t+\epsilon(uv_x+vv_z)]=- p_z\\
 u_x+v_z=0\\
  v=\eta_t+\epsilon u\eta_x  \, \textrm{ and } \,  p=\eta \, \, \textrm{ on }\,
z=1+\epsilon\eta(x,t)\\
 v=0 \, \,
\textrm { on } z=0
 \end{array}
\label{e+bc''} \end{equation} The two important  parameters
$\epsilon$ and $\delta$  that arise in water-waves theories, are
used to define various approximations of the governing equations
and the boundary conditions. The amplitude parameter $\epsilon$ is
associated with the nonlinearity of the wave, so that small
$\epsilon$ implies a nearly-linear wave theory. The shallowness
parameter $\delta$ is associated with the dispersion of the wave,
it measures the deviation of the pressure, in the water below the
wave, away from the hydrostatic pressure distribution.

\section{Variational derivation of the Camassa-Holm shallow water
equation}

In contrast to the KdV equation, which is a classical integrable
model for shallow water waves, the Camassa-Holm equation possesses
not only solutions that are global in time but models also wave
breaking. The only way that singularities can arise in finite time
in a smooth solution is in the form of breaking waves, that is,
the solution remains bounded but its slope becomes unbounded
\cite{c}. Even if a wave breaks, there is a procedure to continue
uniquely the solution after wave breaking \cite{bc}.

There are different approaches which lead to the Camassa-Holm
equation in the shallow water regime. The original derivation of
this equation \cite{holm} consists in making approximations  by
relating $m:=u-u_{xx}$ and $\eta$ in the Green-Naghdi Hamiltonian
system and preserving the momentum part of its Hamiltonian
structure. The Camassa-Holm equation can be also obtained from the
governing equations (\ref{e+bc''}), by the use of  a double
asymptotic expansion, valid for $\epsilon\rightarrow 0$,
$\delta\rightarrow 0$, retaining terms $O(1)$,  $O(\epsilon)$,
$O(\delta^2)$, $O(\epsilon\delta^2)$ (see
for example \cite{johnson1}, \cite{johnson2}
). As a result, a single nonlinear equation for $\eta$ will be
obtained and thus all variables will be expressed through the
solution of this equation. In what follows we will consider a
derivation of the model from the governing equations for the
water-wave problem by a variational approach in the Lagrangian
formalism \cite{adrian1}, \cite{delia}.

One observes that the parameter $\delta$ can be removed from the
system (\ref{e+bc''}) (see \cite{johnson1}), this being equivalent
to use only $h_0$ as the length scale of the problem. In order to
do this,  the non-dimensional variables $x$, $t$ and $v$ from
(\ref{nondim}) are replaced by
 \begin{equation} x\mapsto
\f{\sqrt{\epsilon}}{\delta}x,\quad t\mapsto
\f{\sqrt{\epsilon}}{\delta}t, \quad v\mapsto\f
{\delta}{\sqrt{\epsilon}}v\label{delta}\end{equation} Therefore
the equations in the system (\ref{e+bc''}) are recovered, but with
$\delta^2$ replaced by $\epsilon$ in the second equation of the
system, that is, this equation writes as  \begin{equation}
 \epsilon[v_t+\epsilon(uv_x+vv_z)]=- p_z\label{1}\end{equation}

The classical approximation is the linearized problem obtained by
requiring the amplitude of the surface to be small letting
$\epsilon\rightarrow 0$. The system (\ref{e+bc''}) with the second
equation given by (\ref{1}) becomes linear:
\begin{equation}
\begin{array}{c}
u_t+p_x=0\\  p_z=0\\
 u_x+v_z=0\\
v=\eta_t \,  \textrm{ and } \,  p=\eta \, \, \textrm{ on }\,
z=1\\
 v=0 \, \,
\textrm { on } z=0
\end{array}
\label{small} \end{equation}

\subsection{The case of an irrotational flow}

Firstly, we consider that the fluid flow is irrotational, i.e. the
flow has zero vorticity. Then, in addition to the system
(\ref{e+bc}), we also have the equation
\begin{equation} u_z-v_x=0 \end{equation} Here the velocity components
 $u$ and $v$ are written in the physical (dimensional) variables.
If we non-dimensionalise this equation using (\ref{nondim}),
(\ref{derivate}), we obtain \begin{equation} u_z=\delta^2 v_x
\label{omega}\end{equation} After scaling (\ref{scaling}) and
transformation (\ref{delta}), the equation (\ref{omega}) writes as
\begin{equation} u_z=\epsilon v_x \end{equation} Therefore, in the limit
$\epsilon\rightarrow 0$, we get in addition to the system
(\ref{small}), the equation \begin{equation} u_z=0
\label{5}\end{equation} Thus, under the assumption that the fluid
is irrotational, $u$ is independent of $z$, that is, \be u=u(x,t)
\label{13}\ee
 From the second equation in (\ref{small}), we get that
$p$ also does not depend on $z$. Because $p=\eta(x,t)$ on $z=1$,
we have
\begin{equation} p=\eta(x,t) \, \quad \textrm{ for any } \, \,
0\leq z\leq 1\label{2}\end{equation} Therefore, using the first
equation in (\ref{small}), and taking into account  (\ref{13}), we
obtain  \begin{equation} u=-\int \eta_x(x,t)dt+\mathcal{F}(x)
\label{3'}\end{equation} where $\mathcal{F}$ is an arbitrary
function. Differentiating (\ref{3'}) with respect to $x$ and using
the third equation in (\ref{small}), we get, after an integration
against $z$,
\begin{equation} v=-zu_x=z\left(\int\eta_{xx}(x,t)dt
-\tilde{\mathcal{F}}'(x)\right)\label{4'}\end{equation} We
underline the fact that in our approximation the vertical velocity
component maintains a dependence on the $z$-variable.  In
analyzing the motion of the fluid particles, this means that the
particles below the surface may perform a vertical motion. This is
in agreement with  recent general results obtained in \cite{c2007}
for the periodic steady waves (Stokes waves) and in \cite{CE2} for
the solitary waves, both being solutions of the full Euler
equations.

\noindent Making $z=1$ in (\ref{4'}), and taking into account that
$v=\eta_t$ on $z=1$, we get after a differentiation with respect
to $t$, that $\eta$ has to satisfy the equation
 \begin{equation}
\eta_{tt}-\eta_{xx}=0 \label{eta}\end{equation} The general
solution of this equation is $\eta(x,t)=f(x-t)+g(x+t)$, where $f$
and $g$ are differentiable functions.  It is convenient first to
restrict ourselves to waves which propagate in only one direction,
thus, we choose
\begin{equation} \eta(x,t)=f(x-t) \label{sol}\end{equation}
Therefore, for $u$ and $v$ in (\ref{3'}), (\ref{4'})  we have $
u(x,t)=\eta(x,t)+\mathcal{F}(x),$
$v(x,z,t)=-z\left(\eta_x+\mathcal{F}'(x)\right)$. The condition
$v=\eta_t$ on $z=1$, yields \be
\mathcal{F}(x)=\textrm{const}:=c_0\ee
 Thus, for the irrotational
case the solution of the system (\ref{small}) plus the equation
(\ref{5}), can be written into the form
\begin{equation} \eta(x,t)=f(x-t), \quad u(x,t)=\eta(x,t)+c_0,\quad
v(x,z,t)=-z\eta_x(x,t)=-zu_x\label{solirrot}\end{equation} The
solutions to the shallow water problem are determined by the
evolution of the function $\eta(x,t)$, which represents the
displacement of the free surface from the undisturbed (flat)
state. The solution (\ref{solirrot}) describes the linear,
non-dispersive surface wave.

By consistently neglecting the $\epsilon$ contribution, we will
derive using variational methods in the Lagrangian formalism (see
\cite{adrian1}, \cite{delia}), the equation (\ref{ch}) governing
unidirectional propagation of shallow water waves.

In the Lagrangian picture of a mechanical system, one focuses the
attention on the motion of each individual particle of the
mechanical system. We denote by $M$ the ambient space whose points
are supposed to represent the particles  at $t=0$. A
diffeomorphism of $M$ represents the rearrangement of the
particles with respect to their initial positions. The set of all
diffeomorphisms, denoted Diff$(M)$, can be regarded (at least
formally) as a Lie group. The motion of the mechanical system is
described by a time-dependent family of orientation-preserving
diffeomorphisms $\gamma(t,\cdot)\in$ Diff($M$).
 For a
particle
 initially located at $x$, the velocity at
 time $t$ is  $\gamma_t(t,x)$, this being the material velocity
 used in the Lagrangian description. The spatial velocity, used
 in the Eulerian description, is the flow velocity
 $w(t,X)=\gamma_t(t,x)$ at the location  $X=\gamma(t,x)$ at time
 $t$,
 that is, $w(t,\cdot)=\gamma_t\circ\gamma^{-1}$.
 In the Lagrangian description, the velocity phase
 space is the tangent bundle $T\textrm{Diff}(M)$. In the Eulerian description,
 the spatial velocity is in the tangent space at the identity $Id$ of Diff$(M)$,
 that is, it is an element of the Lie algebra of Diff$(M)$. The
 Lagrangian $\mathcal{L}$ is a scalar function defined on $T\textrm{Diff}(M)$
 and
 the equation of motion is the equation satisfied by a critical
 point of  the action $\mathfrak{a}(\gamma)=\int_0^T\mathcal{L}(\gamma,\gamma_t)dt$ defined
 on all paths  $\{\gamma(t,\cdot),$ $t\in[0,T]\}$ in $\textrm{Diff}(M)$
  having fixed endpoints.

For our problem, we saw in (\ref{solirrot}) that the vertical
component $v$ of the velocity is completely determined by the
horizontal component $u(x,t)$ of the velocity, which is a vector
field on $\mathbf{R}$, that is, it belongs to the Lie algebra of
Diff$(\mathbf{R})$. Thus, in the Lagrangian formalism of our
problem we take $M=\mathbf{R}$
 and add the technical assumption that the
 smooth functions defined on  $\mathbf{R}$ with value in $\mathbf{R}$
  vanish rapidly at $\pm\infty$ together
 with as many derivatives as necessary (see \cite{c} for a possible choice of weighted Sobolev
spaces). For the
 one-dimensional periodic
 motion, one takes $M=\mathbf{S}^1$ the unit circle.
  In what follows we focus on the first
 situation.

\noindent Following Arnold's approach to Euler equations on
diffeomorphism groups \cite{arnold},  the action for our problem
   will be obtained by transporting
  the kinetic energy to all  tangent spaces of
  Diff$(\mathbf{R})$ by means of right translations.
  For small surface elevation, the potential energy is
   negligible compared to the kinetic energy.
   Taking into account (\ref{solirrot}), the kinetic energy on the surface is
\begin{equation} K=\f 1{2}\int_{-\infty}^{\infty}(u^2+v^2)dx=\f
1{2}\int_{-\infty}^{\infty} \left[u^2+(1+\epsilon
\eta)^2u_x^2\right]dx\approx \f 1{2}\int_{-\infty}^{\infty}
\left(u^2+u_x^2\right)dx
   \end{equation}
to the order of our approximation (see \cite{adrian1}). We observe
that if we replace the path $\gamma(t,\cdot)$ by
$\gamma(t,\cdot)\circ\psi(\cdot)$, for a fixed time-independent
$\psi$ in Diff($\mathbf{R}$), then the spatial velocity is
unchanged $\gamma_t\circ \gamma^{-1}$. Transforming $K$ to a
right-invariant Lagrangian, the action on a path
$\gamma(t,\cdot)$, $t\in [0,T]$, in Diff($\mathbf{R}$) is
\begin{equation} \mathfrak{a}(\gamma)= \f
1{2}\int_0^T\int_{-\infty}^{\infty}
\{(\gamma_t\circ\gamma^{-1})^2+[\partial_x(\gamma_t\circ\gamma^{-1})]^2\}dxdt\label{action}\end{equation}
The critical points of the action (\ref{action}) in the space of
paths with fixed endpoints, verify \begin{equation}
\f{d}{d\varepsilon} \mathfrak{a}(\gamma+\varepsilon\varphi)\Big
|_{\varepsilon=0}=0,\label{critic}\end{equation} for every path
$\varphi(t,\cdot)$, $t\in[0,T]$, in $\textrm{Diff}(\mathbf{R})$
with endpoints at zero, that is,
$\varphi(0,\cdot)=0=\varphi(T,\cdot)$ and such that
$\gamma+\varepsilon\varphi$ is a small variation of $\gamma$ on
Diff($\mathbf{R}$). Taking into account (\ref{action}), the
condition (\ref{critic}) becomes \begin{eqnarray}\hspace{-0.65cm}
\int_0^T\int_{-\infty}^{\infty}&&\left\{\left(\gamma_t\circ\gamma^{-1}\right)
\f{d}{d\varepsilon}\Big
|_{\varepsilon=0}\left[(\gamma_t+\varepsilon\varphi_t)\circ(\gamma+\varepsilon\varphi)^{-1}\right
] \right.\nonumber\\
\hspace{-0.65cm}&&\left. +
\partial_x(\gamma_t\circ\gamma^{-1})\f{d}{d\varepsilon}\Big
|_{\varepsilon=0}\left[\partial_x\left((\gamma_t+\varepsilon\varphi_t)\circ(\gamma+\varepsilon\varphi)^{-1}\right)\right]\right\}dxdt=0
\label{6}\end{eqnarray} After calculation (see \cite{adrian1}), we
get \begin{eqnarray} \hspace{-0.65cm}\f{d}{d\varepsilon}\Big
|_{\varepsilon=0}\left[(\gamma_t+\varepsilon\varphi_t)\circ(\gamma+\varepsilon\varphi)^{-1}\right
]=&&\hspace{-0.65cm}\varphi_t\circ\gamma^{-1}-(\varphi\circ\gamma^{-1})\partial_x(\gamma_t\circ\gamma^{-1})
\nonumber\\
=&&\hspace{-0.65cm}\partial_t(\varphi\circ\gamma^{-1})+(\gamma_t\circ\gamma^{-1})\partial_x(\varphi\circ\gamma^{-1})\nonumber\\
&&\hspace{-0.65cm} -
(\varphi\circ\gamma^{-1})\partial_x(\gamma_t\circ\gamma^{-1})
\label{11}\end{eqnarray} and
 \begin{eqnarray}  \hspace{-0.65cm}\f{d}{d\varepsilon}\Big
|_{\varepsilon=0}\left[\partial_x\left((\gamma_t+\varepsilon\varphi)
\circ(\gamma+\varepsilon\varphi)^{-1}\right)\right]
&=&
\partial_x(\varphi_t\circ\gamma^{-1})-\partial_x(\gamma_t\circ\gamma^{-1})\partial_x(\varphi\circ\gamma^{-1})\nonumber\\
&&-(\varphi\circ\gamma^{-1})\partial_x^2(\gamma_t\circ\gamma^{-1})
\nonumber\\
\hspace{-0.65cm}&=&\partial_{tx}(\varphi\circ\gamma^{-1})+(\gamma_t\circ\gamma^{-1})\partial^2_x(\varphi\circ\gamma^{-1})\nonumber\\
&&-(\varphi\circ\gamma^{-1})\partial_x^2(\gamma_t\circ\gamma^{-1})\label{12}\end{eqnarray}
where are used the formulas of the type \begin{equation} \f
{d}{d\varepsilon}\Big
|_{\varepsilon=0}\left[(\gamma+\varepsilon\varphi)^{-1}\right]=-\f{\varphi\circ\gamma^{-1}}{\gamma_x\circ\gamma^{-1}}\label{7}\end{equation}
\begin{equation}
\partial_x(\gamma_t\circ\gamma^{-1})=\f{\gamma_{tx}\circ\gamma^{-1}}{\gamma_x\circ\gamma^{-1}}
\end{equation} \begin{equation}
\partial_t(\varphi\circ\gamma^{-1})=\varphi_t\circ\gamma^{-1}+\left(\varphi_x\circ\gamma^{-1}\right)\partial_t(\gamma^{-1})=
\varphi_t\circ\gamma^{-1}-(\gamma_t\circ\gamma)\partial_x(\varphi\circ\gamma^{-1})
\end{equation}
Thus, denoting $\gamma_t\circ\gamma^{-1}=u$, from (\ref{11}),
(\ref{12}), the condition (\ref{6}) writes as \begin{eqnarray}
\hspace{-0.5cm}\int_0^T\int_{-\infty}^{\infty}&&
\left\{u\left[\partial_t(\varphi\circ\gamma^{-1})+u\partial_x(\varphi\circ\gamma^{-1})
-(\varphi\circ\gamma^{-1})u_x\right]\right.\nonumber\\
\hspace{-0.5cm}&& \left.+ u_x\left[
\partial_{tx}(\varphi\circ\gamma^{-1})+u\partial^2_x(\varphi\circ\gamma^{-1})
-(\varphi\circ\gamma^{-1})u_{xx}\right]\right\}dxdt
=0\label{10}\end{eqnarray} We integrate by parts with respect to
$t$ and $x$ in the above formula, we take into account that
$\varphi$ has endpoints at zero, the smooth functions defined on
$\mathbf{R}$ with values in $\mathbf{R}$, together
 with as many derivatives as necessary, vanish rapidly at $\pm\infty$, and we obtain \begin{equation}
-\int_0^T\int_{-\infty}^\infty(\varphi\circ\gamma^{-1})\left[u_t+3uu_x
-u_{txx}- 2u_xu_{xx}-uu_{xxx}\right]dxdt=0 \end{equation}

\noindent Therefore, we proved:


\begin{theorem}
\textit{
For an irrotational
 unidirectional shallow water flow,   the horizontal velocity component
 of the water $u(x,t)$ satisfies the Camassa-Holm
equation (\ref{ch}) for $\kappa=0$.}

\end{theorem}

Let us see now which equation fulfill the displacement $\eta(x,t)$
of the free surface from the flat state. Taking into account
(\ref{solirrot}), the kinetic energy on the surface is \ba K=\f
1{2}\int_{-\infty}^{\infty}(u^2+v^2)dx&=&\f
1{2}\int_{-\infty}^{\infty} \left[(\eta+c_0)^2+(1+\epsilon
\eta)^2\eta_x^2\right]dx\nonumber\\
&\approx& \f 1{2}\int_{-\infty}^{\infty}
\left[\left(\eta+c_0\right)^2+\eta_x^2\right]dx
   \ea
to the order of our approximation. Transforming $K$ to a
right-invariant Lagrangian, the action on a path
$\Gamma(t,\cdot)$, $t\in [0,T]$, in Diff($\mathbf{R}$) is
\begin{equation} \mathfrak{a}(\Gamma)= \f
1{2}\int_0^T\int_{-\infty}^{\infty}
\{(\Gamma_t\circ\Gamma^{-1}+c_0)^2+[\partial_x(\Gamma_t\circ\Gamma^{-1})]^2\}dxdt\label{action1}\end{equation}
The critical points of the action (\ref{action1}) in the space of
paths with fixed endpoints, verify \begin{equation}
\f{d}{d\varepsilon} \mathfrak{a}(\Gamma+\varepsilon\Phi)\Big
|_{\varepsilon=0}=0,\label{critic1} \end{equation} for every path
$\Phi(t,\cdot)$, $t\in[0,T]$, in $\textrm{Diff}(\mathbf{R})$ with
endpoints at zero, that is, $\Phi(0,\cdot)=0=\Phi(T,\cdot)$ and
such that $\Gamma+\varepsilon\Phi$ is a small variation of
$\Gamma$ on Diff($\mathbf{R}$). Taking into account
(\ref{action1}), the condition (\ref{critic1}) becomes
\begin{eqnarray}\hspace{-0.5cm}
\int_0^T\int_{-\infty}^{\infty}&&\left\{\left(\Gamma_t\circ\Gamma^{-1}+c_0\right)
\f{d}{d\varepsilon}\Big
|_{\varepsilon=0}\left[(\Gamma_t+\varepsilon\Phi_t)\circ(\Gamma+\varepsilon\Phi)^{-1}\right
] \right.\nonumber\\
\hspace{-0.5cm}&&\left. +
\partial_x(\Gamma_t\circ\Gamma^{-1})\f{d}{d\varepsilon}\Big
|_{\varepsilon=0}\left[\partial_x\left((\Gamma_t+\varepsilon\Phi_t)\circ(\Gamma+\varepsilon\Phi)^{-1}\right)\right]\right\}dxdt=0
\label{66}\end{eqnarray} After calculation, denoting
$\Gamma_t\circ\Gamma^{-1}=\eta$, (\ref{66}) writes as
\begin{eqnarray} \hspace{-0.5cm}
\int_0^T\int_{-\infty}^{\infty}&& \left\{(\eta
+c_0)\left[\partial_t(\Phi\circ\Gamma^{-1})+\eta\partial_x(\Phi\circ\Gamma^{-1})
-(\Phi\circ\Gamma^{-1})\eta_x\right]\right.\nonumber\\
\hspace{-0.5cm}&& \left.+ \eta_x\left[
\partial_{tx}(\Phi\circ\Gamma^{-1})+\eta\partial^2_x(\Phi\circ\Gamma^{-1})
-(\Phi\circ\Gamma^{-1})\eta_{xx}\right]\right\}dxdt=0
\label{10'}\end{eqnarray} We integrate by parts with respect to
$t$ and $x$ in the above formula, we take into account that $\Phi$
has endpoints at zero, the smooth functions defined on
$\mathbf{R}$ with values in $\mathbf{R}$, together
 with as many derivatives as necessary, vanish rapidly at $\pm\infty$, and we obtain \begin{equation}
-\int_0^T\int_{-\infty}^\infty(\Phi\circ\Gamma^{-1})\left[\eta_t+3\eta\eta_x
+2c_0\eta_x-\eta_{txx}-
2\eta_x\eta_{xx}-\eta\eta_{xxx}\right]dxdt=0 \end{equation}

 \noindent We thus have:


\begin{theorem}
\textit{For an irrotational
 unidirectional shallow water flow,   the displacement $\eta(x,t)$ of the free
surface from the flat state, satisfies the Camassa-Holm equation
(\ref{ch}) for $\kappa=c_0$}.
\end{theorem}

\subsection{The case of a linear shear flow}

We assume now that the underling flow is rotational with a
constant vorticity $\omega_0$, that is, we are in the case of
linear shear flow. Then, in addition to the system (\ref{e+bc}),
we also have the equation \be
u_z-v_x=\textrm{const}:=\omega_0\label{omega'}\ee If we
non-dimensionalise this equation using (\ref{nondim}),
(\ref{derivate}), we scale using (\ref{scaling}) and we transform
using (\ref{delta}), the equation (\ref{omega'}) writes as \be
u_z=\epsilon v_x +\f{\omega_0\sqrt{gh_0}}{g}\label{5'}\ee
Therefore, for the case of a linear shear flow, in the limit
$\epsilon\rightarrow 0$, we get in addition to the system
(\ref{small}), instead of (\ref{5}), the equation \be
u_z=\f{\omega_0\sqrt{gh_0}}{g} \label{5''}\ee
 The relation
(\ref{2}) remains the same but instead of (\ref{3'}) we have now
\be u=-\int
\eta_x(x,t)dt+\mathcal{F}(x)+\f{\omega_0\sqrt{gh_0}}{g}z
\label{3''}\ee where $\mathcal{F}$ is an arbitrary function. Using
the third equation in (\ref{small}), we get again (\ref{4'}).
Following the same procedure as in the irrotational case,
presented after the relation (\ref{4'}), we obtain the solution of
the system (\ref{small}) plus the equation (\ref{5''}) into the
form \be \eta(x,t)=f(x-t), \quad
u(x,z,t)=\eta(x,t)+\f{\omega_0\sqrt{gh_0}}{g}z+c_0,\quad
v(x,z,t)=-z\eta_x(x,t)\label{solrotconst}\ee

We will derive using the variational methods in Lagrangian
formalism, the equation of the water's free surface $\eta(x,t)$.
Taking into account (\ref{solrotconst}), we observe that
$\eta(x,t)$ determined completely the velocity components $u$ and
$v$. $\eta(x,t)$ can be regarded as a a vector field on
$\mathbf{R}$, that is, it belongs to the Lie algebra of
Diff$(\mathbf{R})$.  The kinetic energy on the surface is \ba
\hspace{-0.3cm}\mathcal{K}=\f
1{2}\int_{-\infty}^{\infty}(u^2+v^2)dx&=&\f 1{2}
\int_{-\infty}^{\infty}
\left\{[\eta+\f{\omega_0\sqrt{gh_0}}{g}(1+\epsilon\eta)+c_0]^2
+ (1+\epsilon \eta)^2\eta_x^2\right\}dx\nonumber\\
&&\approx \f 1{2}\int_{-\infty}^{\infty}
\left[(\eta+\f{\omega_0\sqrt{gh_0}}{g}+c_0)^2+\eta_x^2\right]dx
 \label{action'}  \ea
to the order of our approximation. Transforming $\mathcal{K}$ to a
right-invariant Lagrangian, the action on a path
$\Gamma(t,\cdot)$, $t\in [0,T]$, in Diff($\mathbf{R}$) is \be
\mathfrak{a}(\Gamma)= \f 1{2}\int_0^T\int_{-\infty}^{\infty}
\{(\Gamma_t\circ\Gamma^{-1}+\f{\omega_0\sqrt{gh_0}}{g}+c_0)^2
+[\partial_x(\Gamma_t\circ\Gamma^{-1})]^2\}dxdt\label{action''}\ee
The critical points of the action (\ref{action''}) in the space of
paths with fixed endpoints, verify (\ref{critic1}), for every path
$\Phi(t,\cdot)$, $t\in[0,T]$, in $\textrm{Diff}(\mathbf{R})$ with
endpoints at zero. Taking into account (\ref{action''}), the
condition (\ref{critic1}) becomes \ba \hspace{-0.5cm}
\int_0^T\int_{-\infty}^{\infty}&&\left\{
\left(\Gamma_t\circ\Gamma^{-1}+\f{\omega_0\sqrt{gh_0}}{g}+c_0\right)
\f{d}{d\varepsilon}\Big
|_{\varepsilon=0}\left[(\Gamma_t+\varepsilon\Phi_t)\circ
(\Gamma+\varepsilon\Phi)^{-1}\right
] \right.\nonumber\\
\hspace{-0.5cm} &&\left. +
\partial_x(\Gamma_t\circ\Gamma^{-1})\f{d}{d\varepsilon}\Big
|_{\varepsilon=0}\left[\partial_x\left((\Gamma_t+\varepsilon\Phi)\circ(\Gamma+\varepsilon\Phi)^{-1}\right)\right]\right\}dxdt=0
\label{66'}\ea After calculation, denoting
\hspace{-0.5cm}$\Gamma_t\circ\Gamma^{-1}=\eta$, (\ref{66'}) writes
as \ba \int_0^T\int_{-\infty}^{\infty}&& \left\{(\eta
+\f{\omega_0\sqrt{gh_0}}{g}+c_0)\left[\partial_t(\Phi\circ\Gamma^{-1})+\eta\partial_x(\Phi\circ\Gamma^{-1})
-(\Phi\circ\Gamma^{-1})\eta_x\right]\right.\nonumber\\
\hspace{-0.5cm}&& \left.+ \eta_x\left[
\partial_{tx}(\Phi\circ\Gamma^{-1})+\eta\partial^2_x(\Phi\circ\Gamma^{-1})
-(\Phi\circ\Gamma^{-1})\eta_{xx}\right]\right\}dxdt=0
\label{10''}\ea We integrate by parts with respect to $t$ and $x$
in the above formula, we take into account that $\Phi$ has
endpoints at zero, the smooth functions defined on $\mathbf{R}$
with values in $\mathbf{R}$, together
 with as many derivatives as necessary, vanish rapidly at $\pm\infty$, and we obtain
 \ba \hspace{-0.5cm}
-\int_0^T\int_{-\infty}^\infty(\Phi\circ\Gamma^{-1})\left[\eta_t+3\eta\eta_x
+2\left(\f{\omega_0\sqrt{gh_0}}{g}+c_0\right)\eta_x-\eta_{txx}\right.\nonumber\\
\hspace{0.5cm}-\left.2\eta_x\eta_{xx}-\eta\eta_{xxx}\right]dxdt=0
\ea

\noindent Thus, we get:

\begin{theorem} \textit{For a rotational
 unidirectional shallow water flow with constant vorticity,   the displacement $\eta(x,t)$ of the free
surface from the flat state, satisfies the Camassa-Holm equation
(\ref{ch}) for $\kappa=\f{\omega_0\sqrt{gh_0}}{g}+c_0$}.
\end{theorem}

\subsection{The case of an arbitrary flow}

We consider now an arbitrary underlying flow. From the second
equation in (\ref{small}), and from $p=\eta(x,t)$ on $z=1$, we get
(\ref{2}).
 Therefore, using the
first equation in (\ref{small}), we obtain \be u=-\int
\eta_x(x,t)dt+\mathcal{F}(x,z) \label{3}\ee where $\mathcal{F}$ is
an arbitrary function. Differentiating (\ref{3}) with respect to
$x$ and using the third equation in (\ref{small}), we get, after
an integration against $z$, \be v=z\int\eta_{xx}(x,t)dt
-\mathcal{G}(x,z)+\mathcal{G}(x,0)\label{4}\ee where
$\mathcal{G}_z(x,z)= \mathcal{F}_x(x,z)$ and we have also taken
into account the last condition in the system (\ref{small}).
Making $z=1$ in (\ref{4}), and taking into account that $v=\eta_t$
on $z=1$, we get after a differentiation with respect to $t$, that
$\eta$ has to satisfy the equation (\ref{eta}).  We restrict
ourselves to waves which propagate in only one direction, thus, we
choose $\eta$ into the form (\ref{sol}). Therefore, for $u$ and
$v$ in (\ref{3}), (\ref{4}) we get \be
u(x,z,t)=\eta(x,t)+\mathcal{F}(x,z),\quad
v(x,z,t)=-z\eta_x(x,t)-\mathcal{G}(x,z)+\mathcal{G}(x,0)\label{solarbitrar}\ee
with \be \mathcal{G}_z(x,z)= \mathcal{F}_x(x,z),\quad
\mathcal{G}(x,1)=\mathcal{G}(x,0)\label{fg}\ee
 arbitrary functions.

Using the variational methods in Lagrangian formalism, we will
derive now the equation of the water's free surface $\eta(x,t)$,
which can be regarded as a a vector field on $\mathbf{R}$, that
is, it belongs to the Lie algebra of Diff$(\mathbf{R})$. Taking
into account (\ref{solarbitrar}), and the second condition in
(\ref{fg}), the kinetic energy on the surface will be now \ba
\hspace{-0.5cm}\mathcal{K}=\f
1{2}\int_{-\infty}^{\infty}(u^2+v^2)dx&=&\f 1{2}
\int_{-\infty}^{\infty}
\left\{[\eta+\mathcal{F}(x,(1+\epsilon\eta))]^2 \right.\nonumber\\
\hspace{-0.5cm}&&\left.+\left[-(1+\epsilon
\eta)\eta_x-\mathcal{G}(x,(1+\epsilon\eta))+
\mathcal{G}(x,0)\right]^2\right\}dx\nonumber\\
\hspace{-0.5cm}&\approx& \f 1{2}\int_{-\infty}^{\infty}
\left\{\left[\eta+\mathcal{F}(x,1)\right]^2+\eta_x^2\right\}dx
 \label{action1'}  \ea
to the order of our approximation. Transforming $\mathcal{K}$ to a
right-invariant Lagrangian, the action on a path
$\textrm{\Large{$\Gamma$}}(t,\cdot)$, $t\in [0,T]$, in
Diff($\mathbf{R}$) is \be \mathfrak{a}(\textrm{\Large{$\Gamma$}})=
\f 1{2}\int_0^T\int_{-\infty}^{\infty}
\left\{\left[\textrm{\Large{$\Gamma$}}_t\circ\textrm{\Large{$\Gamma$}}^{-1}+\mathcal{F}(x,1)\right]^2
+[\partial_x(\textrm{\Large{$\Gamma$}}_t\circ\textrm{\Large{$\Gamma$}}^{-1})]^2\right\}dxdt\label{action1''}\ee
The critical points of the action (\ref{action1''}) in the space
of paths with fixed endpoints, verify (\ref{critic1}), for every
path $\textrm{\Large{$\Phi$}}(t,\cdot)$, $t\in[0,T]$, in
$\textrm{Diff}(\mathbf{R})$ with endpoints at zero. Taking into
account (\ref{action1''}), the condition (\ref{critic1}) becomes
\ba \hspace{-0.5cm} \int_0^T\int_{-\infty}^{\infty}&&\left\{
\left[\textrm{\Large{$\Gamma$}}_t\circ\textrm{\Large{$\Gamma$}}^{-1}+
\mathcal{F}(x,1)\right] \f{d}{d\varepsilon}\Big
|_{\varepsilon=0}\left[(\textrm{\Large{$\Gamma$}}_t+\varepsilon\textrm{\Large{$\Phi$}}_t)\circ
(\textrm{\Large{$\Gamma$}}+\varepsilon\textrm{\Large{$\Phi$}})^{-1}\right
] \right.\nonumber\\
\hspace{-0.5cm}&&\left. +
\partial_x(\textrm{\Large{$\Gamma$}}_t\circ\textrm{\Large{$\Gamma$}}^{-1})\f{d}{d\varepsilon}\Big
|_{\varepsilon=0}\left[\partial_x\left((\textrm{\Large{$\Gamma$}}_t+\varepsilon\textrm{\Large{$\Phi$}})\circ(\textrm{\Large{$\Gamma$}}+\varepsilon\textrm{\Large{$\Phi$}})^{-1}\right)\right]\right\}dxdt=0
\label{66''}\ea After calculation, denoting
$\textrm{\Large{$\Gamma$}}_t\circ\textrm{\Large{$\Gamma$}}^{-1}=\eta$,
(\ref{66''}) writes as \ba
\hspace{-0.5cm}\int_0^T\int_{-\infty}^{\infty}&& \left\{\left[\eta
+\mathcal{F}(x,1)\right]\left[\partial_t(\textrm{\Large{$\Phi$}}\circ\textrm{\Large{$\Gamma$}}^{-1})+\eta\partial_x(\textrm{\Large{$\Phi$}}\circ\textrm{\Large{$\Gamma$}}^{-1})
-(\textrm{\Large{$\Phi$}}\circ\textrm{\Large{$\Gamma$}}^{-1})\eta_x\right]\right.\nonumber\\
\hspace{-0.5cm}&& \left.+ \eta_x\left[
\partial_{tx}(\textrm{\Large{$\Phi$}}\circ\textrm{\Large{$\Gamma$}}^{-1})+\eta\partial^2_x(\textrm{\Large{$\Phi$}}\circ\textrm{\Large{$\Gamma$}}^{-1})
-(\textrm{\Large{$\Phi$}}\circ\textrm{\Large{$\Gamma$}}^{-1})\eta_{xx}\right]\right\}dxdt=0
\label{10'''}\ea We integrate by parts with respect to $t$ and $x$
in the above formula, we take into account that
$\textrm{\Large{$\Phi$}}$ has endpoints at zero, the smooth
functions defined on $\mathbf{R}$ with values in $\mathbf{R}$,
together
 with as many derivatives as necessary, vanish rapidly at $\pm\infty$, and we obtain
  \ba
-\int_0^T\int_{-\infty}^\infty(\textrm{\Large{$\Phi$}}
\circ\textrm{\Large{$\Gamma$}}^{-1})\left[\eta_t+3\eta\eta_x
+2\mathcal{F}(x,1)\eta_x+\mathcal{F}_x(x,1)\eta-\eta_{txx}\right.\nonumber\\
\hspace{-0.5cm}\left.-
2\eta_x\eta_{xx}-\eta\eta_{xxx}\right]dxdt=0 \ea

\noindent In conclusion, we proved:

\begin {theorem} \textit{For an arbitrary
 unidirectional shallow water flow,   the displacement $\eta(x,t)$ of the free
surface from the flat state, satisfies a generalized Camassa-Holm
equation}
\begin{equation} F'(x)U+U_t+\left[3U+2F(x)\right]U_x-U_{txx}=2U_xU_{xx}+UU_{xxx}
\label{ch'}\end{equation}\textit{ with $x\in\mathbf{R}$,
$t\in\mathbf{R}$, $U(x,t)\in \mathbf{R}$, $F(x)\in \mathbf{R}$.}
\end{theorem}

\vspace{0.3cm}

\textbf{Acknowledgements}. I would like to thank Prof. A.
Constantin for many interesting and useful discussions on the
subject of water waves and for constructive comments that improved
the initial version of the manuscript.

\medskip

\medskip

\end{document}